\documentclass[fleqn,twoside]{article}
\usepackage[headings]{espcrc2}
\usepackage{amsmath}
\usepackage{graphicx}
\usepackage{amssymb}
\usepackage{flushend}
\usepackage{subfig}
\usepackage[ruled,noend]{algorithm2e}
\usepackage{color}
\usepackage{float}

\let\ab\allowbreak 
\readRCS
$Id: espcrc2.tex,v 1.2 2004/02/24 11:22:11 spepping Exp $
\title{\textbf{Usages of Composition Search Tree in Web Service Composition}}

\author{Lakshmi.H.N\address[hnl]{PhD Scholar, University of Hyderabad, Hyderabad, Contact: hnlakshmi@gmail.com \\},
Hrushikesha Mohanty \address{Professor, University of Hyderabad, Hyderabad.}}

\runtitle{Usages of Composition Search Tree in Web Service Composition}
\runauthor{Lakshmi.H.N, Hrushikesha Mohanty.}

\begin{document}
\begin{abstract}
The increasing availability of web services within an organization and on the Web demands for efficient search  and composition mechanisms to find services satisfying user requirements. Often consumers may be unaware of exact service names that's fixed by service providers. Rather consumers being well aware of their requirements would like to search a service based on their commitments(inputs) and expectations(outputs). Based on this concept we have explored the feasibility of I/O based web service search and composition in our previous work\cite{hnl}. The classical definition of service composition ,i.e one-to-one and onto mapping between input and output sets of composing services,is extended to give rise to three types of service match: Exact,Super and Partial match. Based on matches of all three types, different kinds of compositions are defined: Exact,Super and Collaborative Composition. Process of composition,being a match between inputs and outputs of services,is hastened by making use of information on service dependency that is made available in repository as an one time preprocessed information obtained from services populating the registry. Adopting three schemes for matching for a desired service outputs, the possibility of having different kinds of compositions is demonstrated in form of a Composition Search Tree. As an extension to our previous work, in this paper, we propose the utility of Composition Search Tree for finding compositions of interest like leanest and the shortest depth compositions.
\end{abstract}

\maketitle

\section{INTRODUCTION}
Web Services are self-contained, self-describing, modular applications that can be published, located, and invoked across the Web.As growing number of services are being available, searching the most relevant web service fulfilling the requirements of a user query is indeed challenging.\\

Various approaches can be used for service search, such as,searching in UDDI, Web and Service portals.The techniques for searching web services can be divided into two categories: discovery and composition. By service composition, we mean making of a new service(that does not exist on its own) from existing services.It can be useful when we are looking for a web service for given inputs and desired outputs and there is no single web service satisfying the request\cite{hnl}.\\

Most of the existing algorithms\cite{has,hnt,arp,gek,kwon,lee,zheng} for service composition construct chains of services based on exact matches of input/output parameters to satisfy a given query.However,the making of a chain fails at a point when inputs of a succeeding($I^S$) service does not match exactly with the outputs ($O^P$) of a preceeding service.\\

To alleviate this problem, in \cite{hnl} we propose a Collaborative Composition among such partially matching services for satisfying a desired service outputs, by making match criteria flexible. In addition to exact match we allow partial as well as super match for conditions $O^P \subset I^S$ and $O^P \supset I^S$ respectively.Partial match is of our interest and in \cite{hnl} we have shown the possibility of successful service composition by collaboration of services that make only partial matches. The process of service composition is visualized as a Composition Search Tree \cite{hnl} that arranges services in levels showing the way service compositions can be made to meet the user requirements.Our approach\cite{hnl} results to a scalable implementation for use of RDBMS, a well proven technology. \\

Here, as an extension to our previous work, we propose the utility of Composition Search Tree for finding optimal service compositions. \\

We define two such optimal compositions -
\begin{itemize}
	\item $Leanest Composition$ - A service composition that requires minimum number of web services to satisfy a given query.
	\item $Shortest Depth Composition$ - A service composition satisfying a given query that has minimum depth in the Composition Search Tree.
\end{itemize}

The remainder of this paper is organized as follows. In Section 2 we essay the related work. In Section 3 we give a brief account of our previous work - service composition process using three modes of composition.Also, we explain Composition Search Tree with an example. Section 4 describes the utility of Composition Search Tree. Algorithms for finding Leanest Composition and Shortest Composition are explained in this section. We conclude our work in Section 5.

\section{RELATED WORK} 

In this section, we survey current efforts related to web services composition, built on relational databases, considering input/output parameters of web services.A web service, $ ws $, has typically two sets of parameters from $\left\{ P_i\right\}$ as set of inputs $ws^I$ and set of outputs $ws^O$.Conventionally two services  $ws_i$ and $ws_j$ are said to be composable iff $ws_i^O= ws_j^I$,i.e,$ws_j$ receives all the required inputs from outputs $ws_i$ has\cite{hnl}.\\

Recently, many researchers have utilized techniques in relational database to solve the service composition problem. Lee et al. \cite{kwon,lee} proposed a scalable and efficient web services composition system based on a relation database system.They pre-compute all possible web service compositions,by applying multiple joins on the tables maintained and store them to be used later for web service composition search. PSR system supports web services having single Input and Output parameters. \\

Zheng et al.\cite{zheng} put forward a new storage strategy for web services which can be flexibly extended in relational database. A matching algorithm SMA is proposed that considers the semantic similarity of concepts in parameters based on WordNet.Based on their storage strategy they propose an algorithm:Fast-EP, for searching service composition.\\

The current techniques based on relational database are constrained by usage of multiple joins as well as malady of exact match of input and output parameters.In \cite{hnl} we propose an approach to overcome these difficulties. The criteria for matching is relaxed for partial matching allowing several services to collaborate and provide a desired service. In the current work we further extend the utility of Composition Search Tree for finding optimal service compositions.

\section{I/O MATCH BASED SERVICE COMPOSITION}
In this section, we summarize our previous work\cite{hnl} in which we propose an approach to extend the classical definition of service composition.We first define the problem statement, followed by the various service composition modes proposed, then give a brief explanation of the composition process and finally explain the Composition Search Tree with an example.

\subsection{Problem Statement}
Given a service registry $R= \left\langle P, W \right\rangle $ and a query $ Q =\left\langle  Q^I, Q^O \right\rangle $ ,we need to find set of web services, $ WS \subseteq W$, $WS=\left\{ws_1,ws_2,…,ws_n\right\}$, such that services in WS can be composed to obtain $ Q^O,  \left\{ws_1^O \cup ws_2^O \cup .... \cup ws_n^O \right\}\supseteq Q^O $ ,where 
\begin{itemize}
\item P is a set of parameters,$P=\left\{P_1,P_2,…,P_n\right\}$. 
\item W is a set of web services in the registry,\\ $W=\left\{ws_1,ws_2,....,ws_n\right\}$. 
\item $ ws_i^O $ is a set of output parameters of web service $ws_i$.
\item $\left\{ ws_1^O \cup ws_2^O \cup .... \cup ws_n^O \right\}$ is the union of output parameters of $ws_1^O,ws_2^O,....,ws_n^O$. 
\item $ Q^I 	\subset P $ is set of initial input parameters
\item $ Q^O  \subset  P $ is set of desired output parameters \\
\end{itemize}

\subsection{Service Composition Types}

Given a registry $R= \left\langle P, W \right\rangle $ , any desired set of output parameters,$ D^O\subset P$, can be satisfied by possibly many compositions. To generate such compositions we start by matching the output of services in the registry,$ws_i^O$,with $D^O$ and classify the services on their composability as Exact,Super and Partial. We can readily define two types of compositions - Exact Composition and Super composition, from the Exact Composable and Super Composable services as follows - \\
\begin{enumerate}
\item \textbf{Exact Composition (EC)}: Exact Composition is a composition obtained by using a web service that is Exactly Composable with $ D^O $,i.e. ,$ ws_i^O = D^O $, where $ws_i \in W$.Such a composition would require additional input parameters ($RI^I_{EC}$) than those specified in $Q^I$ given by,
\begin{center}
$RI^I_{EC} = ws_i^I-Q^I $
\end{center}
where $ws_i^I$ is input parameters of web service $ws$.There can be many services in $W$ that are Exactly composable with $ D^O$ and one of them is chosen in each level to be solved further.\\
\item \textbf{Super Composition (SC)}: Super Composition is a composition obtained by using a web service that is Super Composable with $ D^O $, i.e. , $  ws_i^O \supset D^O $, where $ws_i \in W$.The additional input parameters required by such a composition ($RI^I_{SC}$) is given by, 
\begin{center}
$RI^I_{SC} = ws_i^I-Q^I $
\end{center}
where $ws_i^I$ is input parameters of web service $ws$. One of the many services in $W$ that are Super composable with $ D^O $ is chosen at each level to be solved further.\\
\end{enumerate}

Most of the existing algorithms for service composition construct chains of services based on Exact Matches of input/output parameters to satisfy a given query.However,this approach fails when the available services satisfy only a part of the input/output parameters in the given query.This shortcoming motivated us to define a new type of composition - Collaborative Composition, that is obtained by using a set of partial composable services. We define Collaborative Composition as -\\

\textbf{Collaborative Composition (CC)}:  Collaborative Composition is a composition obtained by using a set of partial composable services , $WS$, that can collaboratively satisfy the desired set of output parameters $ D^O $, i.e ,there exists a set of services $WS_{CC}$ , such that 
\begin{center}
$WS_{CC} \subset W$ , $WS_{CC}= \left\{ws_1 ,ws_2,...,ws_n \right\} $ \\~\\
where $ws_i^O \subset D^O$, $\forall ws_i \in WS_{CC}$  \\~\\
and $ \{ ws_1^O \cup ws_2^O \cup ... \cup ws_n^O \} \supseteq D^O $ 
\end{center}
There can be many such service sets that satisfy $ D^O $.The additional input parameters required ($RI^I_{CC}$) to execute the services in $WS_{CC}$ is given by 
\begin{center}
$RI^I_{CC} = WS_{CC}^I-Q^I-WS_F^O $ 
\end{center}
where $WS_{CC}^I$ is collective input parameters required by the set $WS_{CC}$, i.e , 
\begin{center}
$WS_{CC}^I= \left\{ ws_1^I \cup ws_2^I\cup ....\cup ws_n^I \right\}$   
\end{center}
and  $ WS_F$ is a set of services such that 
\begin{center}
$ WS_F \subseteq WS_{CC}$ , $WS_F= \left\{ws_1,ws_2,.. \right\} $ \\~\\
such that $ \forall ws_j \in WS_F  ,ws_j^I  \subseteq Q^I $.
\end{center}

\subsection{Composition Process}
In this section we describe the process of generating service compositions satisfying a given query, proposed in our previous work\cite{hnl}.The steps involved in composition process is as below -
\begin{itemize}
\item \textbf{Search for Matching Services }: The composition process starts with searching for services in the registry whose output parameters match with the required output parameters as specified in the user query($Q^O$).\\
\item \textbf{Classify the Compositions} : The many compositions satisfying $Q^O$ are classified as Exact Composition, Super Composition and Collaborative Composition.We then choose one of the many possible compositions in each type,in each level,and create three child nodes,(Left,Middle and Right), representing each type: Exact, Super and Collaborative Composition,respectively.\\
\item \textbf{Solve for Additional Input Parameters}: In the next level these compositions are solved for the additional input parameters required,to those provided as input parameters in the query($Q^I$).The matching compositions are categorized on their composability mode and one of the compositions of each type is chosen to be solved further in the next level for additional input parameters required.\\
\item \textbf{Repeat process untill all compositions are found}: The process is repeated recursively until the tree explores all compositions satifying the given user query.\\
\end{itemize}

\subsection{Composition Search Tree}
\label{sec:CST}
In order to visualize the composition process and to find all possible compositions that satisfy a given user query we construct a Composition Search Tree\cite{hnl}. The Composition Search Tree supports querying for optimal service compositions such as Leanest Composition and Shortest Depth Composition.\\

\begin{figure}[h]
	\centering
			\caption{BNF of a CST Node }
		\includegraphics[width=\linewidth,height=9cm]{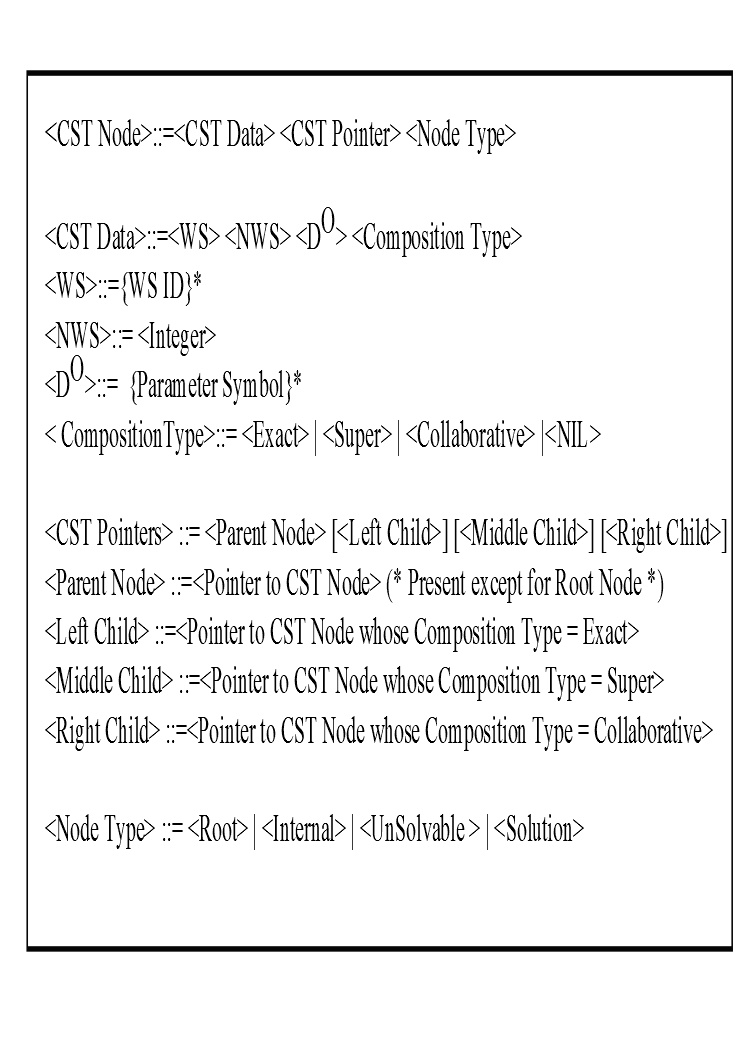}
	\label{CSTNODE}
\end{figure}

The structure of a node in $Composition Search Tree$ is given by Backus Naur Form(BNF) in Fig \ref{CSTNODE}.The abbreviations used in BNF are described in Table \ref{BNF}.

\begin{table}[H]
\centering
	\caption{ Abbreviations used in BNF}
		\begin{tabular}{@{}|l|l|}
		\hline
			\bfseries Abbreviation &	\bfseries Description  \\
	  \hline WS & Set of web services\\
	      &   participating in Composition \\
	 	\hline NWS & Number of web services used \\
		\hline $D^O$ &	Desired set of \\
		    & output parameters \\
		\hline
				\end{tabular}
	\label{BNF}
\end{table}

The $Composition Search Tree$ has 4 types of nodes as described below -
\begin{enumerate}
\item \textbf{Root Node} : A CST node from where the composition process begins,having the following special properties -
	\begin{itemize}
	\item  $  \langle WS \rangle = \emptyset $
	\item  $ \langle NWS \rangle =0 $
	\item  $  \langle D^O\rangle =  Q^O $
	\item  $  \langle Composition Type\rangle = NIL $
	\item  $  \langle Parent Node\rangle = NULL $
	\end{itemize}
	
\item \textbf{Internal Node} : A CST node that represents a composition (Exact,Super or Collaborative) satisfying $D^O$ of its Parent Node.Every internal node of the $Composition Search Tree$ has utmost 3 $Child Nodes$, a $LeftChildNode$ representing $Exact Composition$, a $MiddleChildNode$ representing  $SuperComposition$ and a $RightChildNode$ representing $Collaborative Composition$. Although there may be many compositions in each type : Exact, Super and Collaborative, that satisfy $D^O$ of the Parent Node, we propose to choose one of the compositions in each type for every internal node and hence limit the number of children to three.Note that the Root Node is also an Internal Node with special properties as explained before.

\item \textbf{UnSolvable Node} : A leaf node that cannot be solved further since $D^O$ of such a node does not have matching compositions in $Service Composability Table$. These type of nodes have the following special properties -
	\begin{itemize}
	\item  $\langle Composition Type \rangle =  \langle Exact \rangle |  \langle Super \rangle | \ab \langle Collaborative \rangle $
	\item  $  \langle D^O \rangle = \{ Parameter Symbol \}^* $
	\item  $  \langle Left Child \rangle = NULL $
	\item  $  \langle Middle Child \rangle = NULL $
	\item  $  \langle Right Child \rangle = NULL $
	\end{itemize}

\item \textbf{Solution Node} : A leaf node that need not be solved further since $D^O$ of such a node is $\emptyset $.These type of nodes represent compositions solving the given user query and have the following special properties -
	\begin{itemize}
	\item  $  \langle Composition Type \rangle =  \langle Exact \rangle |  \langle Super \rangle | \ab  \langle Collaborative \rangle$
	\item  $  \langle D^O \rangle = \emptyset $
	\item  $  \langle Left Child \rangle = NULL $
	\item  $  \langle Middle Child \rangle = NULL $
	\item  $ \langle Right Child \rangle = NULL $ \\
	\end{itemize}
\end{enumerate}

Every node of the $Composition Search\ab Tree$ stores the composition that satisfies desired output parameters of its parent node ($WS$),number of web services used for composition($NWS$) and set of additional input parameters required by the composition ($D^O$).\\

Each node in the Composition Search Tree has utmost 3 $Child Nodes$, a $LeftChildNode$ representing $Exact Composition$, a $MiddleChild$ $Node$ representing  $SuperComposition$ and a $RightChildNode$ representing \linebreak $ Collaborative Composition$.\\

\begin{figure*}
	\centering
		\caption{Example CST}
		\includegraphics[width=\linewidth,height={9cm}]{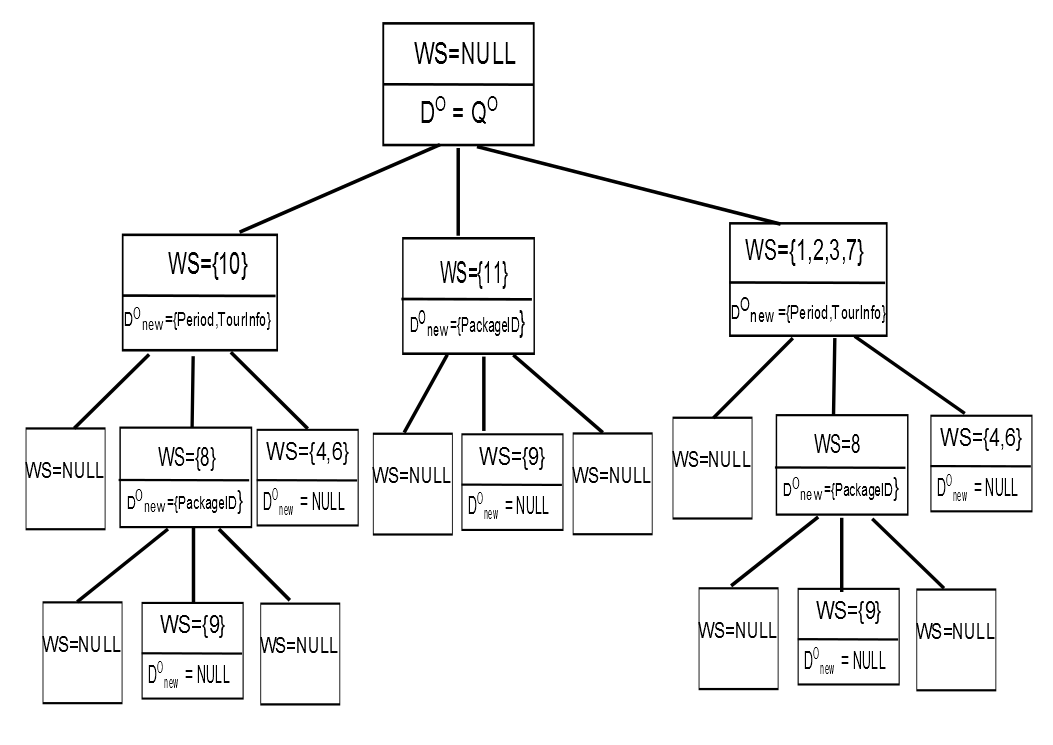}
	\label{fig:CSTFinal}
\end{figure*}

Fig \ref{fig:CSTFinal} depicts Composition Search Tree using the Web services in Table \ref{table:ws} to construct the tree for a query with $Q^I=\left\{Date,City\right\}$ and  $ Q^O =\left\{HotelName,FlightInfo, \linebreak CarType,TourCost\right\}$.\\

\begin{table*}
	\centering
		\caption{ \textbf {Example Web Services}}
		\begin{tabular}{@{} |l|l|l|l|}
		\hline
		\textbf {WS No} & 	\textbf {Service Name} & 	\textbf {Input Parameters} & 	\textbf {Output Parameters} \\
	  \hline	ws1	 & HotelBooking &	Period,City	& HotelName,HotelCost \\
	  \hline	ws2	& AirlineReservation	& Date,City &	FlightInfo,FlightCost\\
		\hline	ws3	& TaxiInfo & Date,City	 &	CarType,TaxiCost \\
		\hline 	ws4	 & DisplayTourInfo &	HotelName, & TourInfo \\
									&      & FlightInfo,CarType & \\
		\hline 	ws5	 & TaxiReservation &	CarType,Date,City &	TaxiCost \\
		\hline 	ws6	 & TourPeriod &	Date,City &	Period \\
		\hline 	ws7	 & TourCost &	TourInfo &	TourCost \\
		\hline 	ws8	 & AgentPackage	& PackageID	& Period,TourInfo \\
		\hline	ws9 &	TourPackages &	Date,City &	PackageID \\
		\hline	ws10	& TourReservation &	Period,TourInfo &	HotelName,FlightInfo,\\
									& & & CarType,TourCost \\
		\hline	ws11	& PackageDetails &	PackageID	& HotelName,Hotelcost,\\
									& & &	FlightInfo,FlightCost,\\
									& & & CarType,TaxiCost,\\
									& & & TourCost\\
		\hline
		\end{tabular}
	\label{table:ws}
\end{table*}

The process for Composition Search Tree construction is given below :
\begin{enumerate}
\item Create a $Root Node$ that has desired output parameters equal to the output parameters specified in the query, i.e $ D^O = Q^O $ ,initialize the number of web services used in composition $NWS$ to $0$ and set of web services participating in composition as empty set, $WS = \emptyset$.
\item $Insert$ the $Root Node$ to $LiveNodesQ$.
\item $Delete$ a $LiveNode$ from $LiveNodesQ$ and set it as the $Current Node$.
\item Find services that match with $D^O$ of the $Current Node$.
\item Classify these services according to their match type.
\item Find different compositions that satisfy $D^O$ from these services based on their match type as -
\begin{enumerate}
\item If a service $ws$ having an Exact Match with $D^O$ is available in registry,create a $LeftChildNode$ for the $Current Node$ ,store $ws$ and update $NWS$ as $NWS = NWS+1$.Calculate the additional input parameters required,to execute $ws$, as $R^I_{EC} = ws^I-Q^I $,where $ws^I$ is input parameters of web service $ws$. $R^I_{EC}$ is the new set of desired output parameters that need to be satisfied, i.e.,$D^O = R^I_{EC}$.If $D^O \neq \emptyset $ then insert the $LeftChildNode$ to $LiveNodesQ$,otherwise mark the $LeftChildNode$ as $SolutionNode$ and $Insert$ a copy of the node to $Solutions$. Make the $Left\ab ChildNode$ point to its $ParentNode$.\\

\item If a service $ws$ having an Super Match with $D^O$ is available in registry,create a $MiddleChild\ab Node$ for $CurrentNode$ ,store $ws$ and update $NWS$ as $NWS = NWS+1$.Calculate the additional input parameters required,to execute $ws$, as $R^I_{RC} = ws^I-Q^I $,where $ws^I$ is input parameters of web service $ws$.$R^I_{SC}$ is the new set of desired output parameters that need to be satisfied,i.e.,$D^O = R^I_{RC}$. If $D^O \neq \emptyset $ then insert the $MiddleChildNode$ to $LiveNodesQ$, otherwise mark the $MiddleChild\ab Node$ as $Solution\ab Node$ and $Insert$ a copy of the node to $Solutions$. Make the $MiddleChildNode$ point to its $Parent\ab Node$.\\

\item Among services that have Partial match find a set of services that can collaboratively satisfy $D^O$.If such a set, $WS$, is available, create a $Right Child\ab Node$ for the $Current Node$,store $WS$ and update $NWS$ as $NWS = NWS+\left|WS\right|$.Calculate the additional input parameters required,if any,to execute the services in $WS$, as $R^I_{CC} = WS^I-Q^I-WS_F^O $ where $WS^I$ is collective input parameters required by the set $WS$, i.e , $WS^I= \left\{ ws_1^I \cup ws_2^I\cup ....\cup ws_n^I \right\}$   and  $ WS_F$ is a set of services such that $ WS_F \subseteq WS ,WS_F= \left\{ws_1,ws_2,.. \right\} \ab | \forall ws_j \in WS_F  ,ws_j^I  \subseteq Q^I $.$R^I_{CC}$ is the new set of desired output parameters that need to be satisfied, i.e.,$D^O = R^I_{CC}$.If $D^O \neq \emptyset $ then insert the $RightChild Node$ to $LiveNodesQ$,otherwise mark the $RightChild Node$ as $Solution Node$ and $Insert$ a copy of the node to $Solutions$. Make the $Right\ab ChildNode$ point to its $ParentNode$.\\

\item If $D^O$ cannot be satisfied by any of the above 3 cases then mark $Current Node$ as $Unsolvable Node$.
\end{enumerate}

\item $Delete$ a $LiveNode$ from $LiveNodesQ$ and set it as the $Current Node$.

\item Find services that match with $WS$ of the $Current Node$.

\item Repeat steps 5 to 8 untill the $LiveNodesQ$ becomes empty.\\~\\
\end{enumerate}

\section{UTILITY OF COMPOSITION SEARCH TREE}
\label{sec:Algms}
As discussed earlier the Composition Search Tree not only finds all possible compositions satifying a given query but can also be utilized for querying for optimal service compositions.In this paper,we define two such optimal service compositions.

\subsection{ \textbf{Leanest Composition}} A service composition that requires minimum number of web services to satisfy a given query is called Leanest Composition.The Leanest Composition is an optimal composition in that it uses least number of services possible for service composition.The procedure for searching a Leanest Composition in Composition Search Tree is given in Algorithm \ref{algo:algo2}.\\

\subsection{ \textbf{Shortest Depth Composition}} A service composition satisfying a given query that has minimum depth in the Composition Search Tree is called Shortest Depth Composition. The Shortest Depth Composition is an optimal composition in that it has least depth in Composition Search Tree.The procedure for searching a Shortest Depth Composition in Composition Search Tree is given in Algorithm \ref{algo:algo3}.\\

\begin{algorithm*}
\SetAlgoNoLine
\LinesNumbered
\DontPrintSemicolon
\KwIn{$Composition Search Tree$}
\KwOut{$Solution Node$ for Leanest Composition}
$CurrentNode=Root Node$ of $CST$ , $Level=0$\;
$CArr$ and $NLCArr$ are arrays of $Child Nodes$\;
\tcp*[l]{$LCN$ is Leanest Composition $Solution Node$}
$LCN=NULL$\;
$minNWS=\infty $ \tcp*[l]{Current minimum number of services}
$Insert$ all $Child Nodes$ of $CurrentNode$ to $CArr$\;
\While{$CArr$ is not empty}
{
	$Level = Level + 1$ \;
	\ForEach{$Child Node$ in $CArr$}
	{	$CN$=$Child Node$\;
		\If{$CN$ is an $Solution Node$}
		{ \tcp*[l] {$NWS$ is number of web services used}
			\If {$Level$ = $NWS$} 
		  { \If { $NWS < minNWS $}
		  	{	\tcp*[l]{$CN$ is the shortest composition}
		  	  $LCN$=$CN$ \nllabel{algo:alg21} \;
					\Return{[$LCN$]} \;
				} 
				\Else
				{	\If{$LCN \neq NULL$}
					{	\tcp*[l]{$LCN$ already present}
						\Return{[$LCN$]} \;
					}
				}
			}
			\Else
			{	\If{ $NWS < minNWS $}
				{	 \tcp*[l]{A shorter composition}  
					$LCN$=$CN$  \nllabel{algo:alg22} \;
				}
			}
		}
		\eIf{$CN$ is an $Unsolvable Node$}
	  { Go To 7\;
	  }
	  ( \tcp*[h]{$CN$ is an $UnResolved Node$})
	  {	$Insert$ all $Child Nodes$ of $CN$ to $NLCArr$\;
	  }
	}
	$CArr$=$NLCArr$ \;
	$CArr$=0\;
}
\If{$LCN \neq NULL$}
{	\Return{[$LCN$]} \;
}
\Else
{	\Return{[$NULL$]} 	\tcp*[l] {Composition Not Found}
}
\caption{Searching Leanest Composition}
\label{algo:algo2}
\end{algorithm*}

\begin{algorithm*}[ht]
\SetAlgoNoLine
\LinesNumbered
\DontPrintSemicolon
\KwIn{$Composition Search Tree$}
\KwOut{$Solution Node$ for Shortest Depth Composition}
\tcp*[l]{Initialization Steps}
$CurrentNode=Root Node$ of $CST$\;
$Level=0$\;
$CArr$ and $NLCArr$ are arrays of $Child Nodes$\;
$Insert$ all $Child Nodes$ of $CurrentNode$ to $CArr$\;
\While{$CArr$ is not empty}
{
	$Level = Level + 1$ \;
	\ForEach{$Child Node$ in $CArr$}
	{	$CN$=$Child Node$\;
		\If{$CN$ is an $Solution Node$}
		{ 				\Return{[$CN$]} \nllabel{algo:alg31} \;
		} 
	  \eIf{$CN$ is an $Unsolvable Node$}
	  { Go To 6\;
	  }
	  ($CN$ is an $UnResolved Node$)
	  {	$Insert$ all $Child Nodes$ of $CN$ to $NLCArr$\;
	  }
	}
	$CArr$=$NLCArr$ \;
	$CArr$=0\;
}
\Return{[$NULL$]} \tcp*[l] {Composition Not Found}
\caption{Searching Shortest Depth Composition}
\label{algo:algo3}
\end{algorithm*}

\subsection{Observations}
The following observations can be made from the Algorithms for Leanest Composition and Shortest Depth Composition -
\begin{itemize}
\item \textbf{Observation 1:} A $Solution Node$ representing $Shortest Depth Composition$ also represents a $Leanest Composition$ if it appears at a $Level$ $i$ that is equal to $NWS$.\\~\\
\textbf{Rationale}: This observation can be reduced from Line number 10 in Algorithm \ref{algo:algo3} and Line number 13 in Algorithm \ref{algo:algo2}. Line number 10 in Algorithm \ref{algo:algo3} always returns the first $Solution Node$ as obtained  in the breadth first search of the Composition Search Tree.If this $Solution Node$ has the property that it appears at a $Level$ $i$ that is equal to $NWS$, then this node will be returned as $Solution Node$ from Line number 13 in Algorithm \ref{algo:algo2}, since this node will have the least $NWS$ among all $Solution Nodes$.

\item \textbf{Observation 2:} A $Solution Node$ represents a $Leanest Composition$ iff there are no other $Solution Nodes$ in the Composition Search Tree that has a lesser $NWS$ than this $Solution Node$.\\~\\
\textbf{Rationale}: This observation can be reduced from Line numbers 13 and 20 in Algorithm \ref{algo:algo2}. These statements search for the $Solution Node$ with the least $NWS$ and hence the algorithm always returns a $Solution Node$ that has the least $NWS$.
\end{itemize}

\section{CONCLUSION}
This paper is an extension to our previous work in \cite{hnl}. In \cite{hnl} the scope of composition is widened defining possibly three modes of service composability: Exact,Partial and Super. Based on composability of all three types and sequencing them differently, $Composition Search Tree$ explores all possible compositions for a given requirement. In the current work, we propose the utility of $Composition Search Tree$ for finding optimal service compositions like $Leanest$  $Composition$ and $Shortest Depth Composition$. \\

The set of web services returned by the algorithms in sections \ref {sec:Algms} and in \cite{hnl} implicitly includes the final composition plan when Exact and Super composition or a combination of the two are involved,given by a chain of service calls from the $Solution Node$ till the $RootNode$.However,a composition plan needs to be derived from the set of services whenever the composition includes Collaborative Composition.In the future work we would like to work on an algorithm that generates a composition plan specifying the order of execution for services participitating in a Collaborative composition.Since our system explores all possible compositions for a given requirement,we would like to include a monitoring component that monitors execution of composition and suggests an alternative composition in case of any service failure.\\

\noindent{\includegraphics[width=1in,height=1.7in,clip,keepaspectratio]{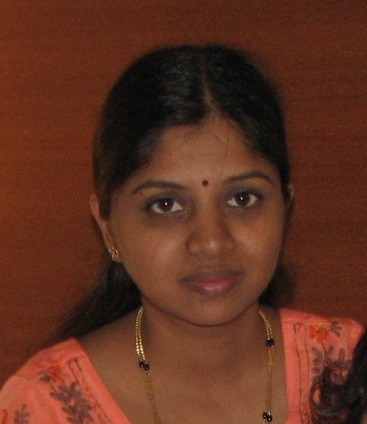}}
\begin{minipage}[b][1in][c]{1.8in}
{\centering{\bf {Lakshmi H N}} is currently a PhD scholar, SCIS(School Of Computer and Information Sciences), University of Hyderabad, Hyderabad. She is working as Associate Professor, CVR College of Engineering, Hyderabad.}\\\\
\end{minipage}
She received her M.S. in Software Systems from BITS Pilani and B.Tech degree from BMS College of Engineering, Bangalore University. Her areas of interest include Web Services, Data Structures, etc.\\\\
\noindent{\includegraphics[width=1in,height=1.7in,clip,keepaspectratio]{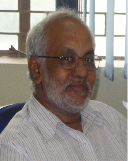}}
\begin{minipage}[b][1in][c]{1.8in}
{\centering{\bf{Hrushikesha Mohanty}} is a Professor in SCIS(School Of Computer and Information Sciences), University of Hyderabad, India. Worked previously in Electronics Corporation of India Limited, Hyderabad.}\\\\
\end{minipage}
Took M.Sc(Mathematics) at Utkal University and completed his Ph.D(Computer Science) from IIT Kharagpur. He has over 85 research publications in refereed International Journals and Conference Proceedings..\\\\

\end{document}